\documentclass[twocolumn,showpacs,aps,prl,nobibnotes,floatfix]{revtex4-1}
\usepackage{graphicx,color}
\usepackage{amsmath,amssymb}

\newcommand{\beq}{\begin{equation}}
\newcommand{\eeq}{\end{equation}}
\newcommand{\beqn}{\begin{eqnarray}}
\newcommand{\eeqn}{\end{eqnarray}}

\newcommand{\apjl}{Astrophys.\ J.\ Lett.\ }
\newcommand{\mnras}{Mon.\ Not.\ R.\ Astron.\ Soc.\ }

\usepackage{dcolumn}
\usepackage{bm}
\usepackage{epsf}

\begin{document}

\title{Binary Black-Hole Mergers in Magnetized Disks:\\Simulations in Full General Relativity}

\author{Brian D. Farris}
\email{bfarris2@illinois.edu}
\affiliation{Department of Physics, University of Illinois at Urbana-Champaign, Urbana, IL~61801}
\author{Roman Gold}
\affiliation{Department of Physics, University of Illinois at Urbana-Champaign, Urbana, IL~61801}
\author{Vasileios Paschalidis}
\affiliation{Department of Physics, University of Illinois at Urbana-Champaign, Urbana, IL~61801}
\author{Zachariah B. Etienne}
\affiliation{Department of Physics, University of Illinois at Urbana-Champaign, Urbana, IL~61801}
\author{Stuart L.\ Shapiro}
\altaffiliation{Also at Department of Astronomy \& NCSA, University of Illinois
at Urbana-Champaign, Urbana, IL 61801}
\affiliation{Department of Physics, University of Illinois at Urbana-Champaign, Urbana, IL~61801}

\bibliographystyle{apsrev4-1}

\begin{abstract}
We present results from the first fully general relativistic,
magnetohydrodynamic (GRMHD) simulations of an equal-mass black hole
binary (BHBH) in a magnetized, circumbinary accretion disk.  We
simulate both the pre and post-decoupling phases of a BHBH-disk system
and both ``cooling'' and ``no-cooling'' gas flows.  Prior to
decoupling, the competition between the binary tidal torques and the
effective viscous torques due to MHD turbulence depletes the disk
interior to the binary orbit. However, it also induces a two-stream
accretion flow and mildly relativistic polar outflows from the
BHs. Following decoupling, but before gas fills the low-density
``hollow'' surrounding the remnant, the accretion rate is reduced,
while there is a prompt electromagnetic (EM) luminosity enhancement
following merger due to shock heating and accretion onto the spinning
BH remnant. This investigation, though preliminary, previews more
detailed GRMHD simulations we plan to perform in anticipation of
future, simultaneous detections of gravitational and EM radiation from
a merging BHBH-disk system.
\end{abstract}

\maketitle

When galaxies merge, massive black hole binaries (BHBHs) are likely to
form in gas rich environments \cite{rodriguez09}. 
These systems constitute unique multi-messenger sources.
The gravitational wave (GW) signals can be detected by future space-based laser
interferometers like eLISA \cite{eLISA12} and Pulsar-Timing-Arrays
\cite{PTA09,tanaka12}. The EM signals provide 
information about MHD accretion onto black holes and can
serve as precursors to GW observations \cite{haiman08}.

If BHBHs are embedded in a gas with negligible angular momentum, the
accretion flow will resemble the Bondi or Bondi-Hoyle-Lyttleton
accretion solutions \cite{farris10,bode11,giacomazzo12}.  When the gas
has intrinsic angular momentum, it will form a circumbinary disk that
accretes via angular momentum transport induced by an effective
viscosity.  The BHBH-disk problem has been studied extensively in
Newtonian gravitation, both
analytically \cite{artymowicz94,milosavljevic05,liu10,kocsis12,rafikov12},
where the emphasis has been on low mass-ratio binaries, and
numerically \cite{artymowicz94,macfadyen08,shi12} where equal-mass
systems have been the focus. The goal is to compute observable EM
``precursor'' and ``aftermath'' radiation that will accompany the GW
signal.  While the vacuum BHBH \cite{pfeiffer12} and accretion onto a
single BH problems in GR are now well developed, simulations of disk
accretion onto BHBHs are still in their
infancy \cite{palenzuela10,farris10,bode11,giacomazzo12,moesta12}.
Vacuum calculations offer an accurate description of the spacetime and
the GWs emitted close to merger in typical cases. Determining the EM
signatures requires a GRMHD computation in this dynamical BHBH
spacetime.

Here we report the first fully GRMHD simulation of a magnetized,
circumbinary BHBH accretion disk. The effective viscosity
driving accretion arises from MHD turbulence triggered by the
magnetorotational instability (MRI) \cite{balbus98}.
This effective viscosity competes with the tidal torques exerted by
the binary, so that a quasi-stationary state is reached prior to
binary-disk decoupling \cite{milosavljevic05,liu10}.  This state has
been simulated both in Newtonian \cite{milosavljevic05} and in
Post-Newtonian \cite{noble12} gravitation. Typically, the
computational domain excludes the region near the BHs and
artificial inner boundary conditions are imposed. Recent Newtonian
studies \cite{roedig12} make clear the importance of imposing the
correct boundary conditions on the flow inside the central disk hollow
and near the BHs, further motivating a treatment in full, dynamical GR
whereby BH horizons can be modeled reliably.

The basic evolution of the system is as follows: For large binary
separations $a$, the inspiral time due to GW emission is much longer
than the viscous time ($t_{\rm GW}\gtrsim t_{\rm vis}$), so that the
disk settles into a quasi-stationary state. For equal-mass BHs, the
binary tidal torques carve out a partial hollow in the disk
\cite{artymowicz94,milosavljevic05,macfadyen08,kocsis12} of radius
$\sim 2a$ and excite spiral density waves throughout the disk, that
dissipate and heat the gas. However, gas can
penetrate the hollow in response to the time-varying tidal torque
\cite{macfadyen08,farris11,kocsis12,noble12}.  At sufficiently small
separations $t_{\rm GW} \lesssim t_{\rm vis}$, and the BHBH
\emph{decouples} from the disk. The disk structure at decoupling
crucially determines its subsequent evolution and the EM emission. GW
emission close to merger leads to mass loss \cite{oneill09,corrales10}
and may induce remnant BH recoil \cite{anderson10}, which give rise to
further characteristic EM signatures.  Here we simulate the system in
two different epochs: (I) The pre-decoupling phase ($t_{\rm GW} >
t_{\rm vis}$) and (II) the post-decoupling phase ($t_{\rm GW} < t_{\rm
  vis}$), including the inspiral and merger of the BHBH. We consider
equal-mass, nonspinning binaries.  While the BH mass scales out, we
are primarily interested in total (ADM) masses $M \gtrsim 10^6
M_\odot$ and low density disks for which the tidally-induced binary
inspiral and the disk self-gravity are negligible.

We use the Illinois numerical relativity code to carry out our
simulations. The code has been extensively tested
\cite{mhd_code_paper,etienne10} and used in our earlier BHBH 
simulations in gaseous media \cite{farris10,farris11}. For details and
equations see \cite{mhd_code_paper,etienne10,etienne12}. The main new
feature concerns our vector potential ($\mathcal{A}_{\mu}=\Phi n_{\mu}
+ A_{\mu}$) formulation for the magnetic induction equation, where
$n^{\mu}$ is the future-directed timelike unit vector normal to a
$t=$const.~slice and $n^{\mu} A_{\mu} = 0$. We introduce a new
generalized Lorenz gauge condition $\nabla_{\mu}\mathcal{A}^{\mu}=\xi
n_{\mu} \mathcal{A}^{\mu}$, where $\xi$ is a parameter (typically
$\xi=4/M$) and $M$ is the total BHBH (ADM) mass. 
This modification results in {\it damped}, traveling EM gauge modes,
preventing spurious B-fields from appearing on refinement
boundaries more strongly than the original, undamped Lorenz
gauge condition \cite{Etienne:2011re}.

The disk initial data represent an equilibrium disk orbiting a single
Schwarzschild BH \cite{chakrabarti85,farris11} with an inner disk edge
at $R_{\rm in} = 18M = 1.8a$, where the specific angular momentum
$\ell_{\rm in}=5.15M$ at $R_{\rm in}$, and a nearly Keplerian rotation
profile parameter $q=1.7$. We adopt a $\Gamma$-law equation of state
with $\Gamma=5/3$, appropriate for a disk composed of an ideal,
nonrelativistic gas.
Magnetic fields are poorly constrained by observations. Thus,
we choose to seed the disk with a weak poloidal B-field as
described in \cite{etienne12}. 
Such initial poloidal B-field configurations 
are widely used in single BH MHD accretion studies (e.g.~\cite{mckinney04}), 
because they facilitate the study of MRI-induced turbulence.
The maximum relative strength of the initial B-field in the equatorial 
plane is $(P_M/P)_{\rm max} = 0.025$. Here $P_M \equiv B^{2}/8\pi$ is magnetic pressure, 
$P$ is gas pressure, and $B^{\mu}$ is the magnetic field measured in the comoving
frame of the fluid. The B-field strength is chosen such that 
it is dynamically unimportant initially, but sufficiently large to capture MRI.

Prior to decoupling we can neglect the slow BHBH inspiral. We model
the spacetime during this epoch by adopting the BHBH metric derived in
the conformal thin-sandwich (CTS) formalism
\cite{cook04}, whereby the spacetime is stationary in the corotating frame (see \cite{farris11} for details). 
The inner part of the disk settles into a quasiequilibrium state on a
``viscous'' time scale
\begin{align}
  \label{tvisc}
  \! \! \frac{t_{\rm vis}}{M}\! =\! \frac{2R_{\rm in}^2}{3\nu M}\! \sim 6500 \left(\frac{R_{\rm in}}{18M}\right)^{3/2} \!\left(\frac{\alpha}{0.13}\right)^{-1}\! \left(\frac{H/R}{0.3}\right)^{-2}\!\!\!,
\end{align}
where $\nu$ is the effective viscosity induced by MHD
turbulence \cite{balbus98}. This viscosity can be fit (approximately)
to an `$\alpha$-disk' law for purposes of analytic estimates.  Here
$R$ is the disk radius, $\nu(R) \equiv (2/3)\alpha
(P/\rho_0) \Omega_K^{-1}\approx (2/3)\alpha (R/M)^{1/2}(H/R)^2 M$, $H$ is
the disk scale height, and we have assumed vertical hydrostatic
equilibrium to derive an approximate relationship between $P/\rho_0$
and $H/R$ (see \cite{shapiro_book_83}).  Equating the viscous time
scale and the GW inspiral time scale yields the decoupling separation
\begin{equation}
  \label{eq:decoupling}
    \frac{a_d}{M} \approx 13
    \left(\frac{\alpha}{0.13}\right)^{-2/5}\left(\frac{H/R}{0.3}\right)^{-4/5},
\end{equation}
where the normalizations give the typical parameters our simulations
obtain for the relaxed state. Note, that for the geometrically thick,
magnetic disks we treat, the expected decoupling radius is an order of
magnitude smaller than typical thin-disk
cases \cite{milosavljevic05,tanaka10}. We thus set our initial binary
separation at $a/M= 10$ (orbital period $2\pi/\Omega=225M$).

We evolve the system using the CTS spacetime for $\sim 45$ binary orbits
(10,000 M) to allow the inner parts of the disk to settle into a
quasistationary state. This epoch (1) models the pre-decoupling phase and (2)
provides realistic, relaxed disk initial data for the post-decoupling
inspiral phase.  
We model the post-decoupling phase by
continuing the GRMHD evolution in the dynamical spacetime
of the inspiraling and merging BHBH binary.
We treat two extreme opposite limiting cases: ``no-cooling'', which
allows for gas heating via shocks induced by tidal torques and MHD
turbulence, and ``cooling'', which removes all the heat generated via
an effective local emissivity $\Lambda$ of the form
$T^{\mu \nu}{}_{;\nu}=-\Lambda u^{\mu}$ as
in \cite{paschalidis11}. Our ``cooling'' case, though artificial,
provides a representative example of the effects of cooling and has
been adopted in previous work (e.g. \cite{penna10,noble12}).  The
particular ``cooling'' prescription we use drives the gas to
isentropic behavior, i.e.  $P/\rho_0^{\Gamma} = const$. The cooling 
timescale is set to a fraction of the local, Keplerian orbital period. Our
simulations resolve the BH horizons and we impose \emph{no inner
boundary conditions}.  In the pre-decoupling phase our grid consists
of a hierarchy of $6$ refinement levels with (coarsest, finest)
resolution of ($5.33M$, $0.16M$) and outer boundary at $250M$. We
resolve the wavelength of the fastest-growing MRI mode ($\lambda_{\rm MRI}$)
by at least $10$ grid points in the bulk of the inner disk.  
Note that resolving $\lambda_{\rm MRI}$ by 10 grid points 
is sufficient to capture the main effects of MRI \cite{shibata06b}.
We add two extra levels centered on each
BH in the post-decoupling phase, increasing the (coarsest, finest)
resolution to ($4M$, $M/32$).
After merger and ringdown we freeze the spacetime evolution, but
continue to evolve the plasma. Equatorial symmetry is imposed
throughout. We normalize results to those for a single BH that we
evolved with the same initial magnetized disk and BH mass equal to 
$M$.

\begin{figure}
  \includegraphics[trim= 0.0 1.3cm 0 0.0cm,clip=True,width=0.35\textwidth]{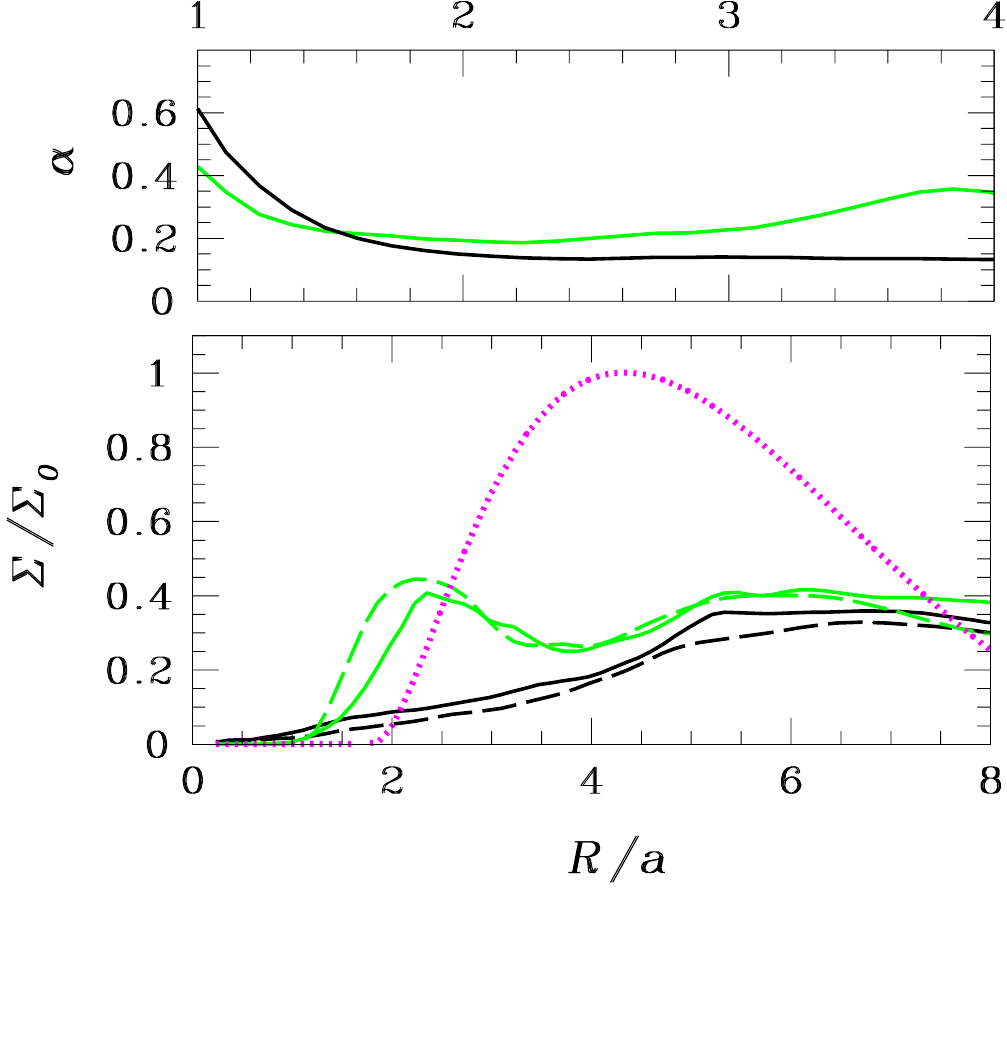}
  \caption{\label{fig:Sigma} 
    The disk $\alpha$ parameter (upper panel) and surface-density $\Sigma$ 
    (lower panel) profiles.
    $\Sigma_0$ is the maximum surface density at $t=0$. 
    Dotted magenta (gray in grayscale) line is the initial data, solid lines are
    at decoupling, and dashed lines are at merger. Black lines are
    from the ``no-cooling'' case, and green (light gray in grayscale) 
    lines are from the ``cooling'' case.   }
\end{figure}

The initial disk (see Fig.~\ref{fig:Sigma}) is not in equilibrium
around the binary as it is perturbed by the binary torques. The torques
lead to spiral density waves in the disk that dissipate and heat 
the gas, puffing up the disk. The gas gains angular momentum
and the surface density profile moves slightly outward.
Magnetic winding converts the poloidal field into one with a large
toroidal component. MRI is induced, resulting in turbulent flow.
After about $20$ binary orbits ($\sim 4-5$ disk orbits at the pressure
maximum) the MRI saturates, driving disk accretion onto the BHBH.
\begin{figure}
  \includegraphics[width=0.4\textwidth]{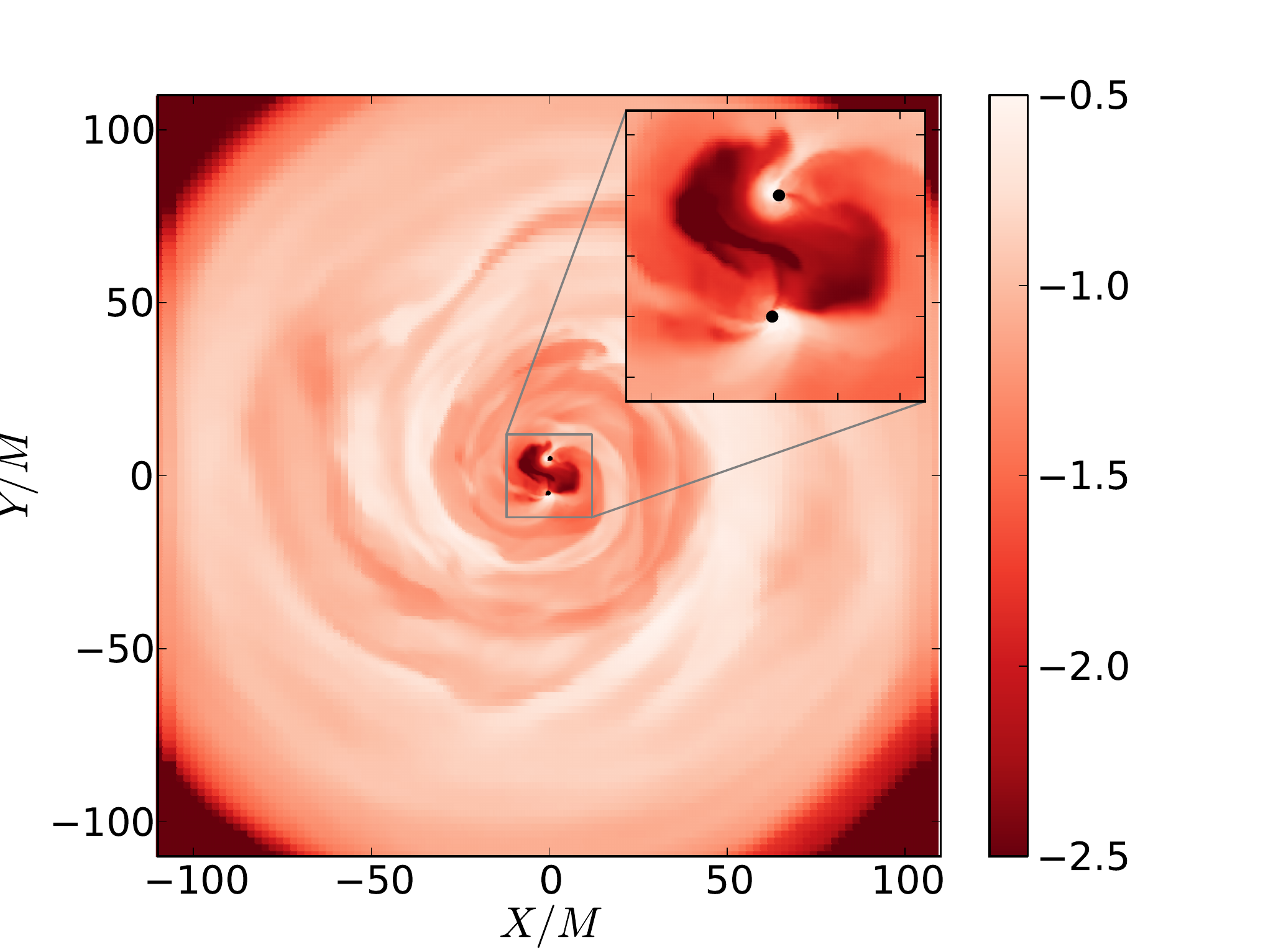}
  \caption{\label{fig:rho_xy} Orbital plane snapshot of rest-mass density $\log(\rho_0/\rho_{0,max})$ 
    from the ``no-cooling'' simulation at
    $t \sim 10000M$ in the relaxed disk, prior to decoupling. The inset zooms in on the region close to the BHs.}
\end{figure}
\begin{figure}
  \includegraphics[width=0.4\textwidth]{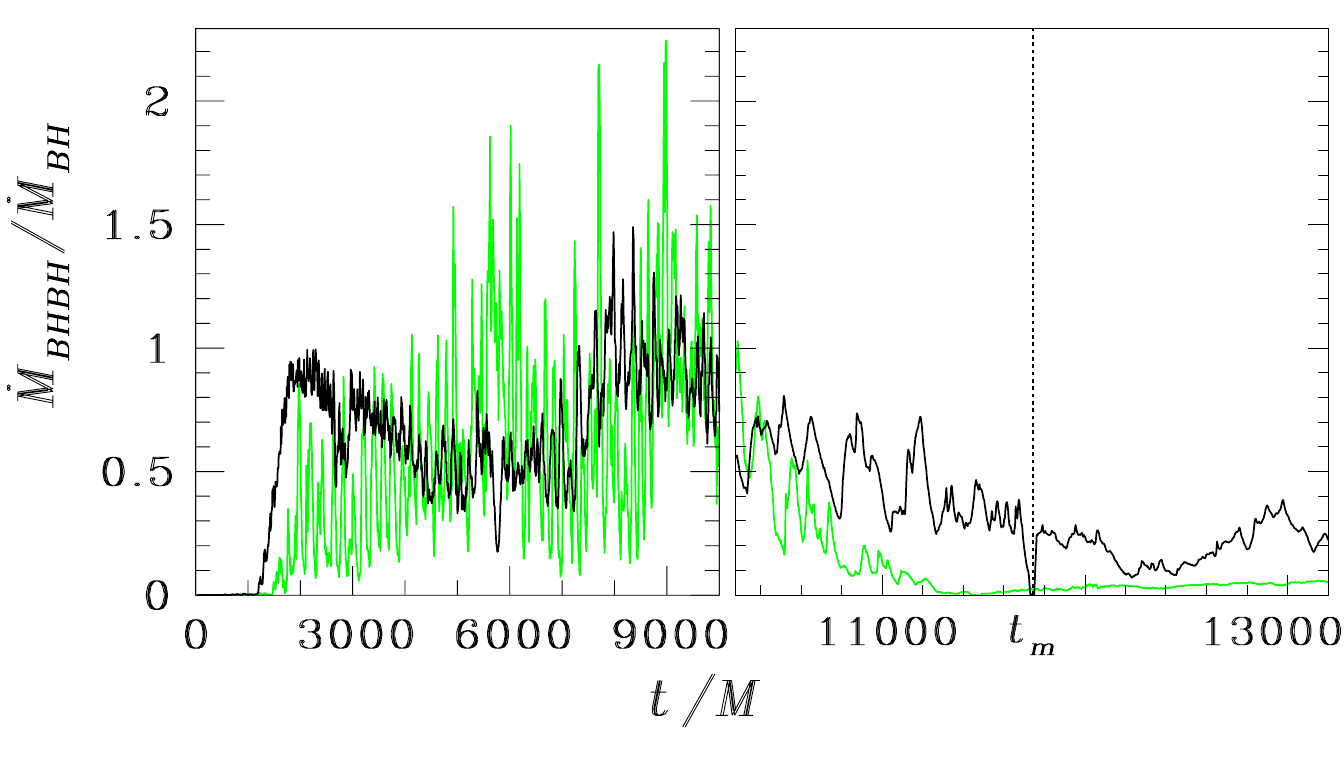}
  \caption{\label{fig:mdot} Time-averaged binary accretion rate $\dot{M}_{BHBH}$, normalized to the average value for a single BH $\dot{M}_{BH}$, versus time. 
   Colors have the same meaning as in Fig.~\ref{fig:Sigma}.
   Left panel: pre-decoupling ($a=$const) phase. Right panel: post-decoupling (inspiral) phase. 
   $\dot{M}_{\rm BH} = 0.45 n_{12} M_8^2 M_\odot \mathrm{yr}^{-1}$ where 
   $M_8 \equiv M/10^8 M_\odot$ and $n_{12} \equiv n/10^{12}\mathrm{cm}^{-3}$ 
   is the initial maximum gas particle number density. Merger occurs at
   time $t_m=11743M$.
}
\end{figure}
In the relaxed disk prior to decoupling we measure a time-averaged
Maxwell-stress as in \cite{kulkarni11} at $20M<R<30M$, and find
$\alpha = 0.13$ for the ``no-cooling'' (see Fig.~\ref{fig:Sigma}) and
$\alpha = 0.2$ for the ``cooling'' case. The magnetic-to-gas-pressure
ratio $1/\beta$ ranges from $0.1$ to $5$ in the bulk of the disk 
(where no numerical fixes or caps are applied).

Cooling influences the global disk structure. In particular, we
observe matter pile-up near the inner disk edge only when there is cooling (see
Fig.~\ref{fig:Sigma}), as has previously been found in
\cite{macfadyen08,liu10,shi12,noble12}. 
The binary maintains a partial hollow in the disk (see
Fig.~\ref{fig:Sigma}) by exerting torques on the plasma, while the
MRI-induced effective viscosity drives matter inward.  Cooling leads
to smaller scale height and lower effective $\nu$, which explains the
enhanced pile-up at $R_{in}$.  We confirm the result in 
\cite{macfadyen08,farris11,shi12,noble12} that non-negligible amounts 
of gas are present inside the cavity.

Accretion occurs predominantly via two spiral density streams inside
the cavity (see Fig.~\ref{fig:rho_xy}). We find that accretion
exhibits an alternating pattern by accreting primarily on one of the
BHs for about half a binary orbit. This is similar to flow features
observed in \cite{shi12}. The
behavior has been attributed to a gradual increase of
disk-eccentricity \cite{macfadyen08}, which weakens one of the two
streams when the BHs pass near the disk-apocenter and strengthens the
other stream at pericenter.

Prior to decoupling, but after an initial transient phase
($5000M \lesssim t \lesssim t_{\rm vis}(R_{\rm in})$), the accretion
rate $\dot M_{\rm BHBH}$ settles to values comparable to those onto a
single BH of mass $M$ (see Fig.~\ref{fig:mdot}). We perform a Fourier
analysis of $\dot{M}_{BHBH}$ and find that the strongest contributions
arise at $2\pi f \sim 2/3 \Omega$ in both of our cases. This is likely
associated with the dominant (2,3) Lindblad
resonance \cite{macfadyen08}.
\begin{figure}
  \includegraphics[width=0.4\textwidth]{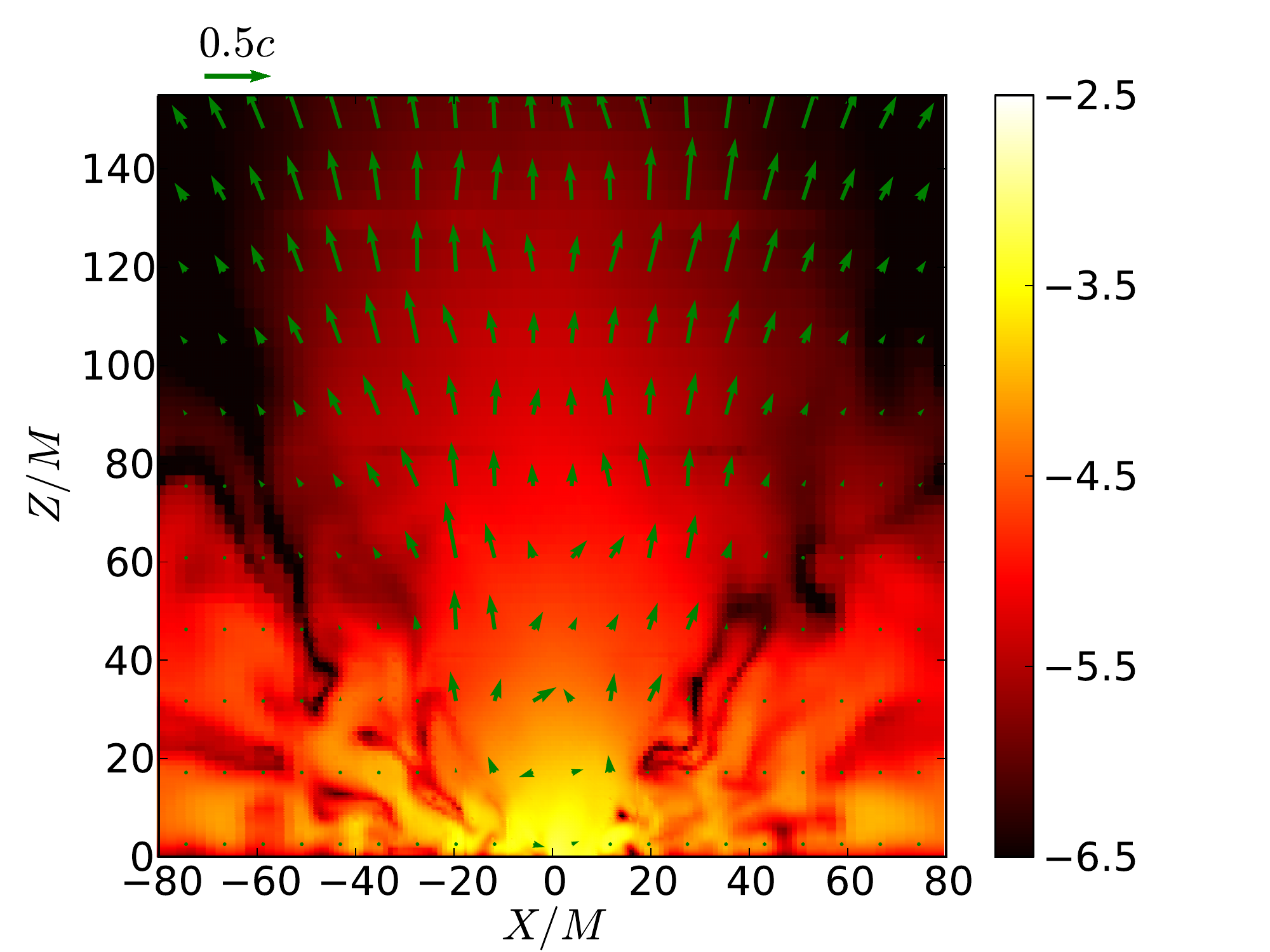}
  \caption{\label{fig:outflow}
  Meridional snapshot of magnetic pressure $\mbox{log}(P_M/\rho_{0,max})$ and fluid velocity vectors at $t \sim 10000M$ for the ``no-cooling'' case.
  }
\end{figure}
During the early inspiral the inward drift of the disk
edge lags behind the binary orbital decay, decreasing $\dot{M}_{BHBH}$. 
In contrast to the magnetic-free case \cite{farris11}, the accretion
streams \emph{remain present} until merger at $t_m$ for the
``no-cooling'' case and up until $t-t_m \gtrsim -400M$ for the
``cooling'' case.
At merger $\dot M_{\rm BHBH}$ decreases gradually to about $30\%$ of
the single, quasi-stationary BH accretion rate, $\dot{M}_{\rm BH}$ (see
Fig.~\ref{fig:mdot}).

The remnant BH settles down via quasi-normal mode ringing to a
Kerr-like BH with mass $M_f\sim 0.95M$ and dimensionless spin
$s=J_f/M_f^2=0.68$. In the ``cooling'' case the inner disk edge
reaches $R_{\rm in}(t_m) \sim 10M$ at merger, while the edge is more
dispersed in the ``no-cooling'' case (see Fig.~\ref{fig:Sigma}).

Prior to decoupling we detect persistent, magnetized, mildly
relativistic ($v \gtrsim 0.5c$) collimated outflows in the polar
regions (see Fig.~\ref{fig:outflow}) in both cases. After merger there
is an increase in the velocities of these outflows. In the "no-cooling"
case the outflows accelerate within a time of $\sim 400M$ to Lorentz factors of
$\Gamma_L \lesssim 4$.  In the "cooling" case this transition takes $\sim
800M$ and the outflow velocities are smaller, $\Gamma_L \lesssim 2$. The
fast outflows persist throughout the postmerger evolution in both cases. 
After merger the effective, turbulent viscous torque will
cause the gas to refill the cavity and accrete on the merger remnant.
Thus, brightening crucially depends on the surface density at
decoupling. If there is only a small pile-up and the majority of gas
lies at radii $R \gtrsim 40M$, then a significant brightening will
take $t_{\rm vis}(R \gtrsim 40M) > \mathcal{O}(10^4M)$ following
merger.  Using Eq.~(\ref{tvisc}) we estimate $t_{\rm vis}(R=40M) \gtrsim
22000M$ ($\alpha \sim 0.13,H/R \sim 0.3$) for the ``no-cooling'' case
and $t_{\rm vis}(R=14M)\gtrsim 6800M$ ($\alpha \sim 0.27,H/R=0.17$) for the
``cooling'' case, respectively. We observe the surface density profile
diffusing inward, but we have not followed the evolution for this long.
\begin{figure}
  \includegraphics[width=0.4\textwidth]{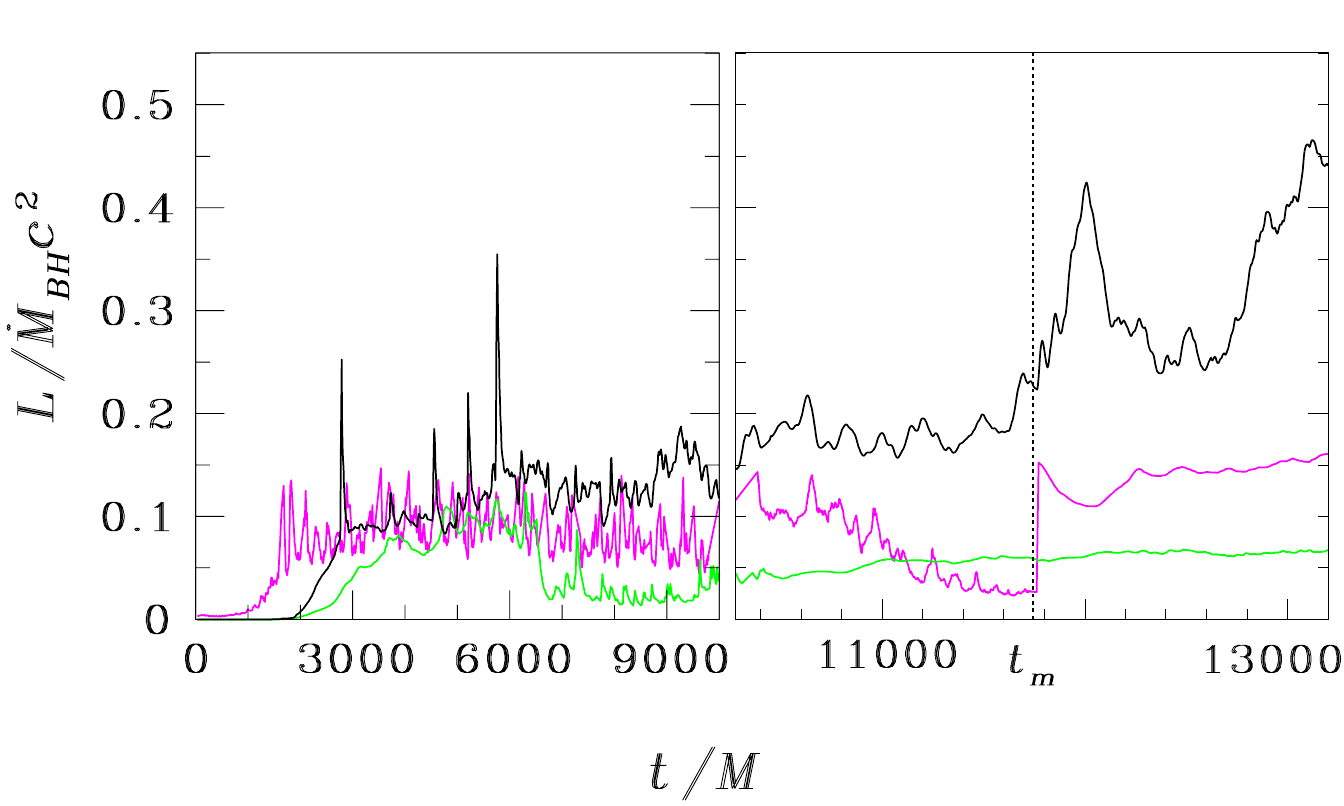}
  \caption{\label{fig:poynting} The Poynting luminosity $L_{\rm EM}$ 
    measured at $R=100M$ as a function of time [green (light gray in grayscale) line for 
    ``cooling'', black line for ``no-cooling'']. The cooling luminosity 
    $L_{cool}$ from the ``cooling'' case [magenta (gray in grayscale) line].
    $\dot{M}_{\rm BH} c^2 = 2.58 \cdot 10^{46} \mathrm{erg} 
    \mathrm{s}^{-1} n_{12} M_8^2$.
}
\end{figure}

According to \cite{BodePhinney2007} an initially quasi-equilibrium, 
hollowed out disk around a non-recoiling remnant BH whose mass is 
suddenly reduced due to energy radiated away in GWs will be 
shock heated as the inner disk edge retracts supersonically, 
leading to prompt EM emission. However, pseudo-Newtonian MHD and 
HD simulations \cite{oneill09} find that shock heating from BH mass loss occurs only if 
the energy loss in GWs $\Delta E_{\rm GW}/M$ is greater than the disk 
half-thickness $H/R$. This limits this process to thin disks only. 
Moreover, \cite{oneill09} shows that for thick disks instead, 
the total luminosity should \emph{decrease} as the disk inner 
edge retracts (a result found to apply to GRHD thick disks 
in \cite{megevand09} and to Newtonian thin disks in \cite{corrales10}). 
Based on this picture the luminosity arising from our thick disk 
($ H/R ~\sim 0.3 $) should not increase, because  
$\Delta E_{\rm GW}/M \sim 0.05 \ll H/R $. Instead, we should observe 
a decline in the total luminosity according to this criterion.

By contrast, we find (Fig.~\ref{fig:poynting}) 
a sudden \emph{increase} in the total luminosity [Poynting luminosity 
($L_{\rm EM} \equiv -\int T^{r}{}_{t}^\mathrm{(EM)}\sqrt{-g}dS$)
plus luminosity from disk cooling 
($L_{cool} \equiv \int \Lambda u_t \sqrt{-g} d^3x$)] after merger. 
This enhancement originates from shocked gas in the immediate 
vicinity of the binary and from accretion onto the spinning BH remnant, 
which taps the BH rotational energy, boosting jet outflows along 
the BH spin axis. This provides a new picture for prompt EM signals 
arising from {\it thick}, relativistic, circumbinary MHD disks 
following the merger.
We find that the outward flux of kinetic energy 
is much smaller in both cases. The total energy efficiency 
$\epsilon \equiv L/\dot{M}_{BH}c^2$ increases to 
$\epsilon = 0.25$ at merger, 
before settling down to $\sim 0.1$ at late times \cite{mckinney04}.  

The Poynting luminosity is presumably reprocessed at larger distance
from the remnant. The characteristic frequencies of the total emitted
EM radiation will depend on the BH masses, disk densities and dominant
cooling mechanisms. 
Relaxing equatorial symmetry will allow advection through the
equator to occur. This in turn will likely enhance the poloidal B-fields
and thereby increase the wavelength of the fastest growing MRI mode,
leading to improved modeling of MRI effects \cite{etienne12b}. 
In addition, the MHD outflow structure may be sensitive to the
initial B-field topology, although the accretion flows and field
saturation levels are probably not \cite{beckwith08}.
We plan to investigate these issues, along with
other precursor (e.g. twin jets) and afterglow effects, different mass
ratios, and different BH spins more thoroughly in future work.

We thank Y.T.~Liu for useful discussions.  We are grateful to
H. Pfeiffer for providing CTS initial data for the BHBH spacetime
metric. This letter was supported in part by NSF Grants No.~PHY-0963136 and
No.~AST-1002667, and NASA Grants No.~NNX11AE11G and No.~NNX09AO64H at the University
of Illinois at Urbana-Champaign. This work used the Extreme Science
and Engineering Discovery Environment (XSEDE), which is supported by
NSF Grant No.~OCI-1053575.

\bibliography{ms}
\end{document}